\documentclass[runningheads]{llncs}
\usepackage{graphicx}
\usepackage{todonotes}
\usepackage{url}

\begin{document}

\newcommand{\repeatthanks}{\textsuperscript{\thefootnote}}
\title{A Case for VR Briefings: Comparing Communication in Daily Audio and VR Mission Control in a Simulated Lunar Mission}
\titlerunning{A Case for VR Briefings}
%
\author{Kinga Skorupska\thanks{The first two authors contributed equally to this study.}\inst{1,2}\orcidID{0000-0002-9005-0348} \and
Maciej~Grzeszczuk\repeatthanks \inst{1}\orcidID{0000-0002-9840-3398} \and
Anna~Jaskulska\inst{1,3}\orcidID{0000-0002-2539-3934} \and
Monika~Kornacka\inst{1,2}\orcidID{0000-0003-2737-9236} \and
Grzegorz~Pochwatko\inst{4,3}\orcidID{0000-0001-8548-6916}\and
Wiesław~Kopeć \inst{1,2,3}\orcidID{0000-0001-9132-4171}
} 
\authorrunning{Grzeszczuk and Skorupska et al.}

\institute{Polish-Japanese Academy of Information Technology\\
\and
SWPS University of Social Sciences and Humanities \and
Kobo Association \and
Institute of Psychology, Polish Academy of Sciences}

\maketitle              
\begin{abstract}
Alpha-XR Mission conducted by XR Lab PJAIT focused on research related to individual and crew well-being and participatory team collaboration in ICE (isolated, confined and extreme) conditions. In this two-week mission within an analog space habitat, collaboration, objective execution and leisure was facilitated and studied by virtual reality (VR) tools. The mission commander and first officer, both experienced with virtual reality, took part in daily briefings with mission control. In the first week the briefings were voice-only conducted via a channel on Discord. During the following week last briefings were conducted in VR, using Horizon Workrooms. This qualitative pilot study employing participatory observation revealed that VR facilitates communication, especially on complex problems and experiences, providing the sense of emotional connection and shared understanding, that may be lacking in audio calls. The study points to the need to further explore VR-facilitated communication in high-stake environments as it may improve relationships, well-being, and communication outcomes.

\keywords{remote collaboration \and virtual reality \and communication}

\end{abstract}
\section{Rationale}

In high-stake environments, where status needs to be reported carefully and often the mode of communication plays a crucial role. The choice between different communication spaces: audio, video or VR, may affect the quality of communication, relationships and well-being of the participants. Especially in the context of manned space flight communication between the crew and Mission Control (MC) this choice may greatly impact goal completion, as problems encountered in space are analysed, simulated and solved by specialized teams on Earth. 

To explore this topic we devised a pilot study to observe how the quality of briefings may change if they are moved from the audio space to virtual reality. In a two-week simulated Lunar mission we had the commander and the first officer take part in daily briefings with MC. In the first week the briefings were voice-only conducted via a locked channel on Discord. During the following week last briefings were conducted in VR, using Horizon Workrooms. This qualitative preliminary study employing participatory observation gathers insights, considerations and future research directions by exploring such research questions as:

\begin{enumerate}
\item To what extend does the technology, audio transmission and VR constitute a barrier to communication? (explored in section \ref{technology})
\item What are the differences between the impressions of participants of remote audio and VR briefings? (explored in section \ref{impressions})
\item What is the potential of using VR-communication to improve the well-being of people experiencing ICE (isolated, confined and extreme) conditions? (explored in section \ref{mental})
\end{enumerate}

\subsection{The role and forms of communication}

Communication is a form of exchanging information, thoughts and feelings. It can be job related, inter-team coordination, or used by mission control to manage the performance of crew members \cite{bondaruk2017}. Communication can be an important factor in isolation missions, stabilizing the well-being as a carrier of relations with friends, family or a wider group of associates \cite{vanhove2015}. The level of copresence during mediated communication (usually higher in rich media like VR) influences well-being \cite{swidrak2020copresence}. Natural communication usually takes place in a multimodal setting, integrating many senses such as sight, hearing and touch. The sender naturally uses those to encode the message not only in the words or the way they are uttered, but also in their posture, gestures or emotions that are conveyed through facial expressions \cite{acarturk2021}. Traditional media - like email, telephone or videoconference - negatively affect the process by filtering out some information carriers \cite{wickman2008}, which matters, as we sometimes send contradictory signals (e.g. when hurt, we say ,,it's fine''). In Mehrabian's study, only 7\% of the actual message is attributed to words, 38\% is vocal and 55\% is the body language \cite{mehrabian2017nonverbal}. Moreover, communication between diverse stakeholders who experience different conditions, such as being under stress in a quick response scenario, is also challenging. While there exist communication protocols and best practices, a shared mental model based on prior training and relevant experience in the personal, interpersonal, and institutional dimensions is necessary to foster effective communication and collaboration, facilitating relationship-building and shared understanding \cite{volcanoComms_2017}.

\subsection{ICE conditions in Human Space Flight}

The subject of human exploration of outer space raises an increased interest in the safety of astronauts staying in the hostile environment of a space mission. Unique factors associated with it, such as microgravity, altered day-night cycle, sense of danger, isolation and confinement, affect human behavior, performance and well-being, but their long-term effects are still not well understood. ICE (Isolated, Confined and Extreme) environments are therefore used for their exploration, combining presence of various type of stressors to uncover compound effects they might have on the human being \cite{zivi2020}.
Experience shows that the impact of ICE conditions is often underestimated by both the crew and the support staff \cite{bishop2004}. The planned return of man to the Moon, and in the longer term, plans to place a man on the surface of Mars, prompt scientists to look for ways to obtain the data, knowledge and experience necessary to prepare the crews for the mission. Building analog habitats allows to simulate selected mission conditions in an artificially created environment. Depending on the project, the emphasis is on study of life support, food self-sufficiency or isolation issues. \cite{schlacht2016}

\subsection{VR for Collaboration and Work}

The use of Virtual Reality (VR) as an environment to meet together, whether for entertainment such as computer games or for professional purposes such as working on a joint project is called "Social VR" \cite{mutterlein2018}. Trials showed that, if the process is properly prepared, e.g. the proper division of roles and competences is ensured so that each team member knows what tasks are ahead and how to use the tools provided, the use of VR is well received and positively influences the sense of community in the team \cite{heinonen2022}. A virtual multi-user experience (VMUE) can also be used as relatively low cost means of simulating the concurrent activities of crews interacting with a simulated work environment. VR applications include training, as well as research on team behavior and interaction during simulated task loads. It can be used to simulate unfamiliar environments, low-gravity conditions or complex equipment maintenance, enabling true collaboration, not just dry procedural training \cite{silva2017,mastro2021}. ESA is exploring the potential of VR and Augmented Reality (AR) also to increase the safety of teams working during missions - the EdcAR project could reduce the number of mistakes during procedures, such as the need for medical assistance when an on-board medic needs help \cite{esa2017}. The recent COVID-19 pandemic gave also an opportunity to explore mental health issues exposed by isolation \cite{chouker2020} and the potential use of VR as a tool to reduce anxiety, depression, perceived stress, and hopelessness while increasing social connectedness \cite{riva2020}.

\section{Method}

\subsection{Simulated Lunar Mission and its Crew}

The research team engaged in a simulated Alpha-XR Mission conducted by XR Lab PJAIT between the 27th of April 2022 and 13th of May 2022 at an analog space habitat. It was initiated by a 3-day training, followed by two weeks of confinement in ICE (isolated, confined and extreme) conditions. The participatory research conducted during the mission related to individual and crew well-being and team collaboration. It was facilitated and studied with the use of virtual reality tools, questionnaires and ethnographic methods. There were five study participants, who, as crew members consented to the research conducted as part of the Alpha-XR mission, and other research ongoing in the analog space habitat. All of the crew members were quite comfortable with information and communication technologies. The two commanders who took part in this particular study were very familiar with VR, having worked with it in a professional setting before and owning an Oculus headset.

\subsection{Briefings with Mission Control}
Every day of the mission, from the flight, lunar landing and the following EVA missions to return to Earth started with a remote briefing with the Mission Control team on Earth. The briefings would start at 9:00 AM and last about an hour, depending on how many things had to be discussed and who took charge of the meeting (either mission control members or the two commanders). Mission Control briefings started with the commanders reporting on the status of the habitat, including the water and energy consumption, graywater levels, food supply. Information about the crew was also included, such as calorie intake, health concerns, any conflicts or changes noticed as well as morale and performance evaluations. Next, the status of the tasks planned for the previous day was discussed, as well as the tasks for the day. Most tasks were related to the planned EVAs\footnote{EVA, that is Extravehicular Activities, were planned for every second day, and they constituted a major part of the simulation, with the exploration of the lunar surface, building up the SWAMP temporary moon base environment (adding sensors and the Internet connection) or testing different analog astronaut suits, such as the BORP suit as well as documenting the lunar surface.}, habitat upgrades and related engineering tasks as well as the XR research that was being conducted as the main aim of the mission.

\begin{figure}[h!]
\centering
\includegraphics[width=0.8\linewidth]{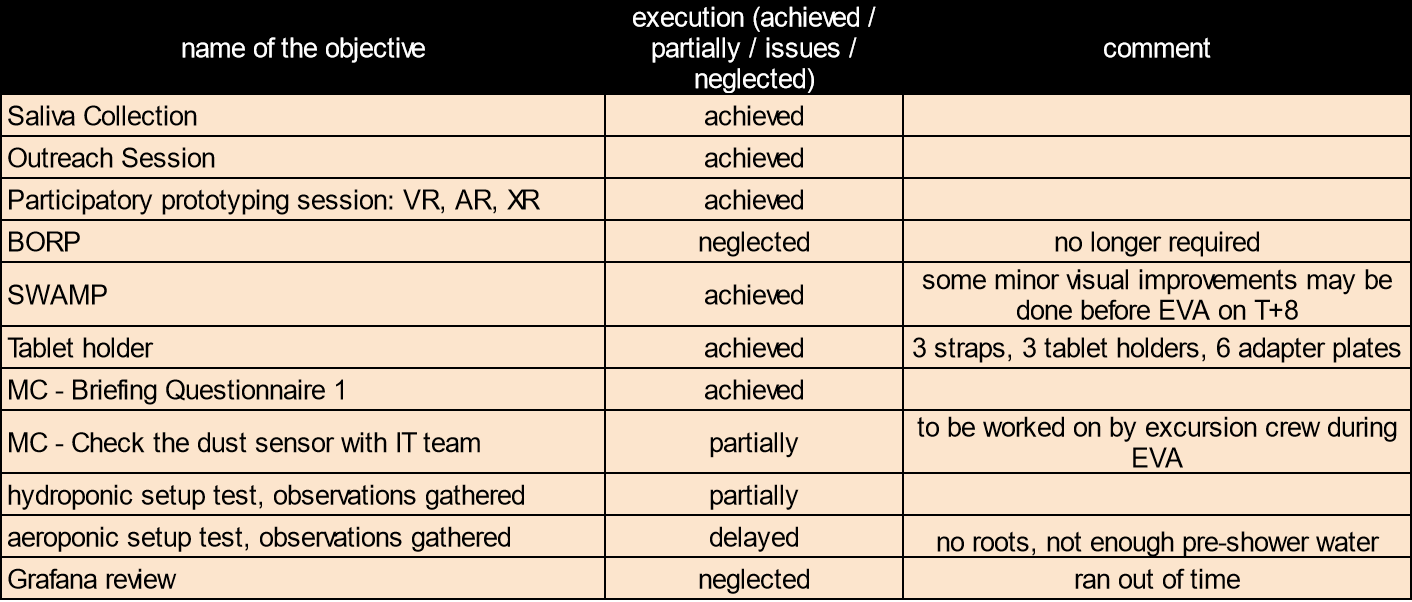}
\caption{Overview of task statuses from Day 6, as discussed on Day 7 of the mission} \label{tasks}
\end{figure}

\subsubsection{Audio Briefings}

Audio briefings were held on a dedicated channel on the habitat's Discord server, and the Commanders participated in them using either their personal laptops, or the personal mission tablets given to them at the start of the mission as well as personal headphones. During the audio briefings all of the participants looked at a shared Google Spreadsheets file, which had a sheet dedicated to the activities and habitat and crew statuses of each day. Additionally, there was an EVA file, with all the objectives and tasks related to the next EVA.

\begin{figure}[h!]
\centering
\includegraphics[width=\linewidth]{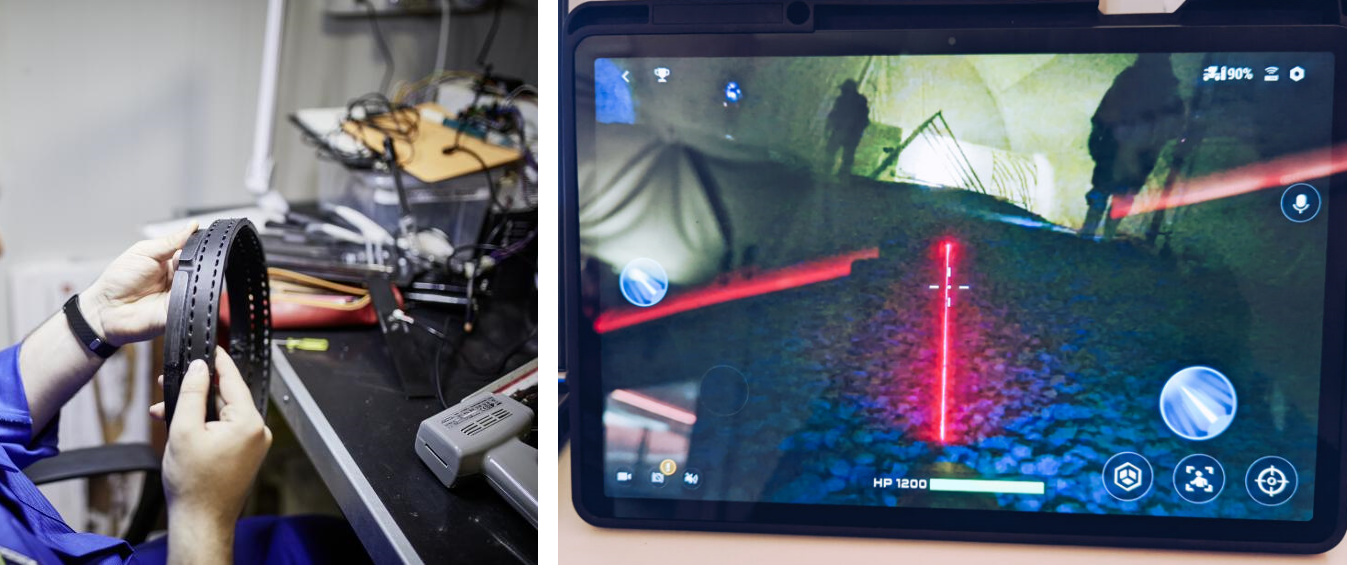}
\caption{Example activities reported during daily briefings. On the left engineering of the BORP suit collar, on the right an EVA mission in progress viewed from a lunar rover camera by HabCom, that is the two crewmates staying behind as support. Photos adapted from originals by Suzanne Baumann (left) and Rafał Masłyk (right); with permission. } \label{eva}
\end{figure}

\subsubsection{VR Briefings}

Two commanders and one member of MC participated in the VR briefings using Oculus 2 VR standalone headsets, on the first day in the same room, and the following days in separate rooms. Additionally, one member of MC connected to these briefings using a desktop application without using their webcam. The VR briefings were held using the Meta Horizon Workrooms and after the first meeting the whiteboard functionality was utilized to discuss tasks and their progress: an image of a relevant snippet of the Google Spreadsheets file was captured and uploaded to the Horizon Workrooms and pulled up to the whiteboard.

\subsection{Briefing Evaluation Tools}
All data was gathered using a timed survey built with Movisens: at 10:00 AM, after each briefing the commanders had an alarm on their tablets, which took them to the installed Movisens app where they jumped right into filling out a brief questionnaire. If this alarm was missed, then they received another prompt at a later time. Most questionnaires were submitted a few minutes after 10:00. The questionnaires consisted of a query on the mode of participating in the meeting (Audio or VR), technical difficulties, discomfort and three longer questions asking to rate the experience on sliding visual analog bipolar scales, expressed underneath by ratings from 0 to 100. 

\begin{itemize}
\item Meeting satisfaction was measured with a scale developed by Rogelberg et al. \cite{Roge2010} after some adjustments. We have adjusted the labels of the scales to instead of "stimulating / boring" read "engaging / non engaging:, instead of "enjoyable  / unpleasant" to "pleasant / unpleasant" and instead of "satisfying  / annoying" to "meets expectations / doesn't meet expectations". 
\item Social presence questions are inspired by Nowak and Biocca \cite{NowakPresence_2003} and Bailenson, Blascovich, Beall and Loomis \cite{bailenson2003interpersonal} (but see also Swidrak, Pochwatko and Matejuk \cite{swidrak2020copresence}); social interaction by Nezlek, Imbrie and Shean \cite{nezlek_1995}.
\item The measures related to the mental state recorded after each meeting were based on Kashdan et al. \cite{Kashdan2013DistinguishingHA} and Ciarocco, Twenge, Muraven, and Tice \cite{ciarocco2007measuring}. 
\end{itemize}

Additionally, meeting notes and impressions after the briefings were recorded in a journal by the executive commander and the crew kept written journals of their experience, as part of an autoethnographic project of another crewmate.

\begin{figure}[h!]
\centering
\includegraphics[width=\linewidth]{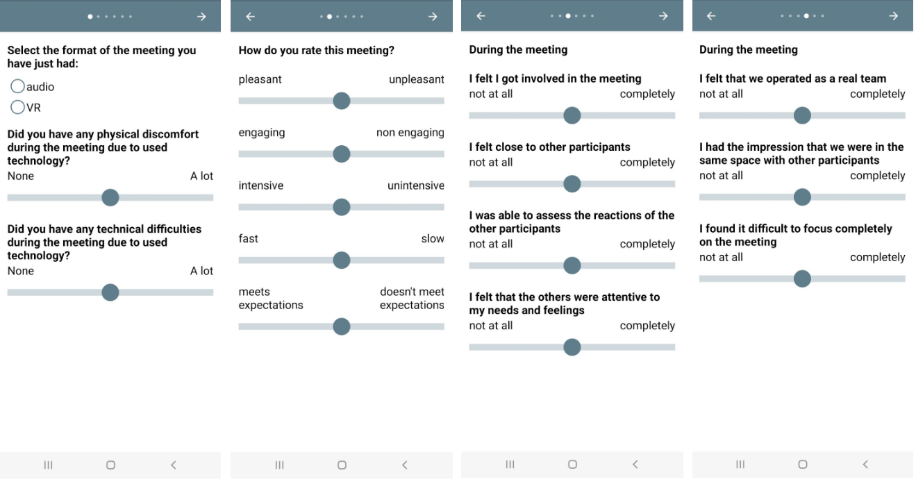}
\caption{Movisens application, showing the first four screens (out of six) of the post-briefing questionnaire employing visual analog bipolar scales. The visible questions are derived from the Meeting Satisfaction scale \cite{Roge2010} } \label{during}
\end{figure}
\vskip  -1em

\section{Results and Discussion}

The quantitative data consists of ratings from two participants of 3 audio meetings (6-8.05.2022), followed by three VR meetings (9-12.05.2022). The data for 5.05 and 10.05 was excluded as it was incomplete. The data was gathered after a week of regular audio meetings, so that participants could get used to holding daily briefings in their default mode before introducing first the questionnaire, after each audio briefing, to establish a baseline, and then the VR meeting mode.

\subsection{Technology and Overall Rating} \label{technology}

\begin{figure}[h!]
\centering
\includegraphics[width=1\linewidth]{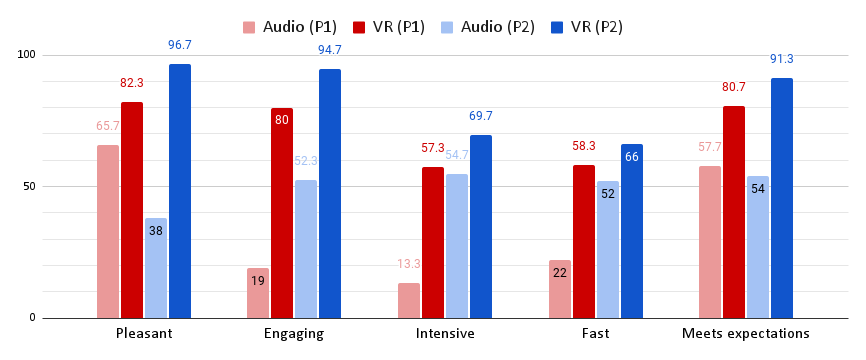}
\caption{Rating averages of the audio and VR meetings} \label{rate}
\end{figure}

Both audio and VR were rated as comfortable and free of difficulties with VR causing slightly more physical discomfort. Journal notes on the first VR meeting read: \textit{There were two short freezes, but other than that it was enough to conduct the meeting. (...) We need to do it in two different rooms, because it is confusing to hear each other twice.} The overall rating of the experience in answer to the question "How do you rate this meeting?" is strongly in favour of VR, as it was rated as nearly double as pleasant, engaging, intensive and fast when compared with audio meetings (see Fig. \ref{rate}). 

Overall, the briefings held in VR were rated at 86/100 vs 55.9/100 for Audio, on whether they have met expectations.

\subsection{Impressions during the Meeting} \label{impressions}

The impressions of the audio meetings recorded in the journal were that the meetings felt slow and not engaging, as if the Mission Control Staff was experiencing time slower, it felt as if they even talked slower. This is recorded as the cause that the commanders were likely to lose focus during audio briefings and drift off, or start doing other tasks meanwhile. The impressions of the perception of time by MC and their subjective reported effect on the commanders may in part be attributed to the issues studied by Kansas et al, who found that with increased quantity of communication, there is an increasing chance of astronauts displacing the problems they experience onto ground control and experiencing more negative effects of ICE \cite{Kanas2001}. Meanwhile, journal notes from the first VR meeting read:
\textit{It was awesome! Much better than the audio meetings we’ve had. The pacing was better (it was same time but more filled with information) and it was nice to share the same space, because we could use gestures, especially to explain some engineering that we’re doing (picture + gestures for VKK helmet) and gestures for the BORP on how it has broken at the collar attachments and the lights and the gloves. We also weren’t as frustrated and shared more, as we felt we can be understood and gauge the others’ impressions of what we are saying. The commander said he never wants to have audio briefings again :)} Especially the use of gestures, enabled by VR communication, seems to be highly beneficial for communication quality while explaining engineering issues or expressing emotions. Similar impressions are visible in survey results in Fig. \ref{during}, as the participants felt as if they participated in face-to-face meetings, sharing the same space with others, who were attentive to their needs.

\begin{figure}[h!]
\centering
\includegraphics[width=\linewidth]{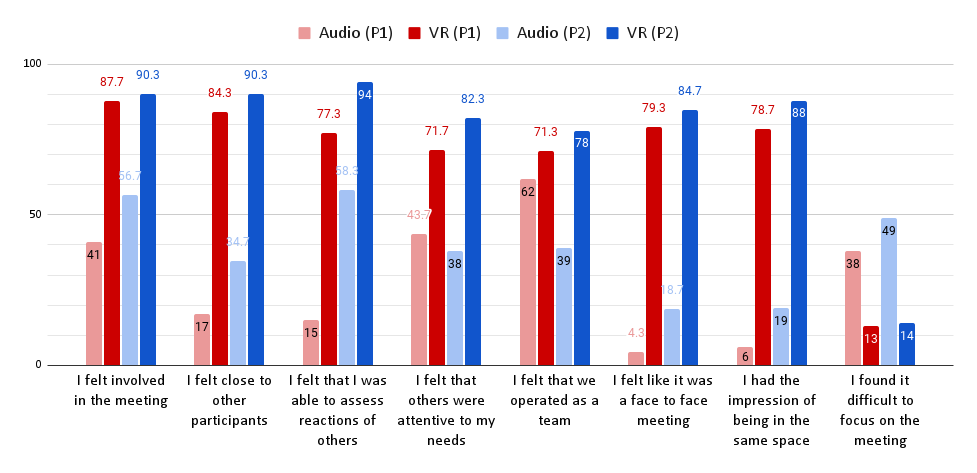}
\caption{Average impressions of the Audio and VR Meetings from P1 and P2} \label{during}
\end{figure}
\vskip  -1em

\subsection{Mental state after each meeting} \label{mental}

\begin{figure}[h!]
\centering
\includegraphics[width=\linewidth]{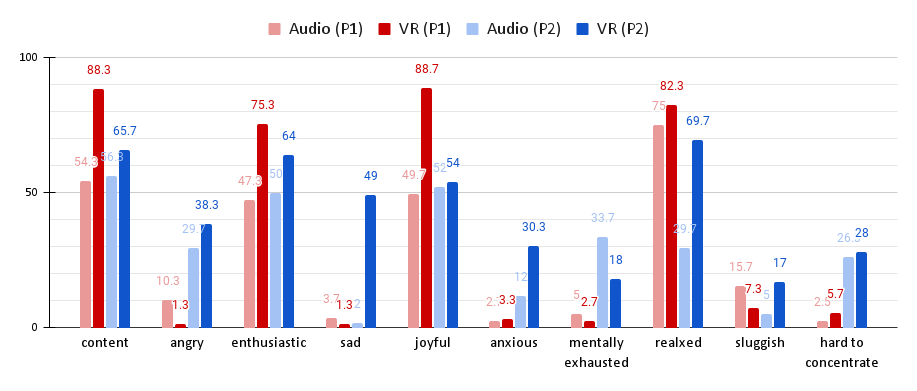}
\caption{Average mental states of P1 and P2 after each type of the meetings} \label{after}
\end{figure}

Mental states were mostly more positive after the use of VR, as shown in Fig. \ref{after}. The only exception here was being "sad" and "anxious" for P2, however that may be attributed either to the slowly approaching ending of the mission, which also coincided with VR meetings or other, unrelated, circumstances. However, on other occasions the whole crew mentioned that as the mission was drawing to an end they felt progressively sadder. Overall, after having used VR for briefings the participants were a bit more content, enthusiastic, joyful, and relaxed, while after audio briefings they were slightly more mentally exhausted. An interesting research direction is the effect of VR on mood and feelings about one's cognitive performance. In this study, the crew mentioned increasing problems as more time in confinement passed, as indicted by this journal entry: \textit{We have cognitive decline that we don’t recognize, because it sneaks up on us. Astronauts have it too - why does that happen though? (...) This is actually a bad problem because it frustrates people. We have trouble learning new stuff. Forget e.g. chords, have trouble learning and explaining new games, have trouble remembering where stuff is either physically or on the drive.}

\vskip  -1em

\section{Conclusions and Future Work}

This pilot study points to possible positive effects of enriching audio-only communication with VR, especially in stressful settings, requiring frequent reports and demanding high understanding between participants. VR facilitates communication, especially when explaining complex problems and experiences and it provides the sense of connection and empathy, that may be lacking when it comes to audio calls. VR-meetings forced participants to fully focus on them and to be in the moment, despite their perceived cognitive decline and the stress of meeting daily goals. This resulted in feeling of being in a shared space, not only physically, but also in terms of common understanding and emotional connection. Future work ought to explore the effect of substituting other modes of communication, either audio or video, with VR in longer, larger studies.

\section*{Acknowledgments}
We would like to thank the many people and institutions gathered together by the distributed Living Lab Kobo and HASE Research Group (Human Aspects in Science and Engineering) for their support of this research. This research is part of ALPHA-XR -- a new long-term proposal for a transdisciplinary research framework programme that fits in the Strategic Innovation Area of Human and Robotic Exploration also known as Terrae Novae programme. 
%
%
%

\bibliographystyle{splncs04}
\bibliography{bibliography}

\end{document}